\documentclass[11pt, a4paper]{article}
\usepackage[margin=1.0in]{geometry}
\usepackage{authblk}
\usepackage{jcappub}
\usepackage{lineno}
\usepackage{mathtools}
\usepackage{physics}
\usepackage{graphicx}					  
\graphicspath{{figs/}}

\usepackage{amssymb}
\usepackage{amsmath}
\usepackage{siunitx}
\usepackage{enumitem}
\usepackage[textwidth=80pt]{todonotes}
\usepackage{caption}
\usepackage{subcaption}
\usepackage{physics}

\usepackage[perpage, para]{footmisc}

\numberwithin{equation}{section}

\newcommand{\lrfrac}[2]{\left( \frac{#1}{#2} \right)} 
\newcommand{\mpl}{\ensuremath{m_\text{pl}}}
\newcommand{\teva}{t_{\text{eva}}}
\newcommand{\teq}{t_{\text{eq}}}
\newcommand{\aeq}{a_\text{eq}}
\newcommand{\keq}{k_\text{eq}}
\newcommand{\MHeq}{M_{H,\text{eq}}}
\newcommand{\GammaM}{\Gamma_\text{merging}}
\newcommand{\GammaE}{\Gamma_\text{cl, evap}}
\newcommand{\Trh}{T_\text{rh}}

\title{Accretion Effects on Primordial Black Hole Reheating Constraints}
\date{\today}
\author{Chenhuan Wang}
\emailAdd{cwang1@uni-bonn.de}
\affiliation{Bethe Center for Theoretical Physics and Physikalisches Institut, Universit{\"a}t Bonn, \\
Nussallee 12, 53115 Bonn, Germany}
\abstract{
    In this work, we study the effects of accretion on the primordial black hole (PBH) reheating scenario. PBHs could form from primordial fluctuations. If they have the right mass and abundance, they could dominate the Universe and complete the reheating entirely through Hawking radiation. We find accretion effects on the BH can not only increase the BH mass, but also prolong such early matter domination. The consequence of the accretion is further investigated using isocurvature induced gravitational waves (GWs), which are generated right after the sudden evaporation of the BHs from the oscillation of the gravitational potential. Big Bang nucleosynthesis limits on the energy density of the GWs put important constraints on the PBH domination scenario. Inclusion of accretion shifts such constraints significantly towards smaller formation mass and smaller initial abundance. Furthermore, the PBH could undergo mergers leading to extended mass functions. We find similar shifts in the allowed parameters with the inclusion of accretion for the merger constraint. We find the constraints from isocurvature GWs typically stronger than the constraints from mergers.
}
\begin{document}
\maketitle 

% \listoftodos

\section{Introduction}
Cosmic inflation is a compelling framework to generate primordial fluctuations. These fluctuations serve as the ``seeds'' for anisotropies observed in the cosmic microwave background (CMB), giving rise to structure formation in the late Universe \cite{planck10, gorbunovIntroductionTheoryEarly2011, baumannTASILecturesInflation2012}. Furthermore, this framework is able to solve several problems in the standard cosmology \cite{gorbunovIntroductionTheoryEarly2011, baumannTASILecturesInflation2012, kolb-turner}. Usually, it is assumed that the inflaton is coupled to additional matter field(s) and coherent oscillations after inflation transfer the energy density of inflaton field to such matter fields. Then, the ``graceful exit'' is complete and the usual hot big bang Universe is recovered \cite{kolb-turner, baumannTASILecturesInflation2012, gorbunovIntroductionTheoryEarly2011}. 

Inflation also allows a plethora of additional phenomena, depending on the exact model. If the primordial fluctuations are enhanced at a certain scale, some fraction of the cosmic horizon could collapse into primordial black holes (PBHs) during radiation domination, once such fluctuations enter the horizon \cite{carrPrimordialBlackHoles2020}. If such black holes (BHs) are abundant enough, the Universe could be dominated by BHs and enter an epoch of early matter domination (eMD), since the black holes at large scales behave just like non-relativistic particles and their energy density redshifts slower than radiation. We denote this era as the PBH domination era. The final reheating then occurs due to the Hawking evaporation of these BHs. Such BHs have to be light enough ($m \lesssim 10^{9}\si{\gram}$) in order to evaporate and reheating the Universe before Big Bang nucleosynthesis (BBN), to not spoil its predictions. This PBH reheating scenario has gained attention lately, see e.g. Refs.~\cite{papanikolaouGravitationalWavesUniverse2021, inomataGravitationalWaveProduction2020, domenechGravitationalWaveConstraints2021, domenechFormationEvaporationInduced2025, haquePrimordialBlackHole2023}. 

Such light black holes, which evaporate away before BBN, are poorly constrained \cite{carrPrimordialBlackHoles2026, carrConstraintsPrimordialBlack2021}. Here we focus on two potential constraints on such scenario: through isocurvature induced gravitational waves (GWs) and through a merger-induced extended mass function. During such early matter domination, the density contrast grows linearly with the scale factor and the gravitational potential doesn't decay \cite{domenechGravitationalWaveConstraints2021, gorbunovIntroductionTheoryEarly2011}. After the sudden evaporation of the BHs, the Universe gets reheated and the gravitational potential starts to oscillate due to the radiation pressure. These oscillations then source gravitational waves at second order \cite{domenechGravitationalWaveConstraints2021, domenechFormationEvaporationInduced2025, papanikolaouGravitationalWavesUniverse2021, inomataGravitationalWaveProduction2020, inomataEnhancementGravitationalWaves2019}. Gravitational waves would contribute to the radiation energy density, which is constrained by BBN at the few percent level \cite{planck6}. As we shall see, this puts important constraints on the formation mass and the initial abundance of BHs. In Ref.~\cite{domenechFormationEvaporationInduced2025}, the epoch prior to the BH formation is not assumed to be radiation domination, but with a general equation of state parameter $\omega \in (0, 1]$. The bounds from GWs depend on the parameter $\omega$ as well. Due to the linear growth of the density contrast in eMD, the density contrast could exceed unity in which case clusters of BHs could form \cite{holstClusteringRunawayMerging2024}. In such dense environments, BHs are more likely to pass by each other and lose energy by GWs and eventually merge. This could produce (much) heavier BHs than their formation mass. An initially monochromatic mass function of BHs gets modified into an extended mass function. If the mass of these merged BHs exceeds $10^{9}\si{\gram}$ and they are abundant enough, their evaporation would have been observed \cite{carrPrimordialBlackHoles2026}. This puts an additional constraint on the scenario.

There is another important effect on black holes after the formation: accretion. The black hole mass increases by absorbing the surrounding fluid. Ref.~\cite{bondiSphericallySymmetricalAccretion1952} considered such spherical symmetric accretion on a point mass. This has been further extended in a general relativistic context in Ref.~\cite{kalitaRevisitingPBHAccretion2025}. We would like to point out that significant accretion has been observed in the numerical simulation of BH formation \cite{escrivaSimulationPrimordialBlack2020}. In general, BH mass could increase by a factor of $\order{1}$. Here, we aim to check the consequences of accretion on the two aforementioned constraints on the PBH reheating scenario.

In Section \ref{sec:background}, the equations for the homogeneous background quantities are solved and analytical estimates are derived. The isocurvature induced gravitational waves and the associated constraints are discussed in Section \ref{sec:iso-gw}. Cluster formation and merging are treated in Section \ref{sec:merger}. We present our conclusion in Section \ref{sec:conc}. Throughout this work, we use the reduced Planck mass $\mpl = (8\pi G)^{-1/2} \approx \SI{2.435e18}{\giga\eV} \approx \SI{4.341e-6}{\g}$. We acknowledge the usage of the following programs: \cite{danischMakiejlFlexibleHighperformance2021, rackauckasDifferentialEquationsjlPerformantFeatureRich2017}. 
The numerical code is available at \href{https://github.com/not-physicist/PBHReheatingAccretion}{Github}.

\section{PBH Reheating}
\label{sec:background}
\subsection{Background equations}
We consider a Universe described by a general equation of state parameter $\omega \in (0, 1] $ upon PBH formation\footnote{Matter domination at formation is explicitly excluded, since the BHs cannot dominate the Universe.}. This can be realized, for example, during reheating with the T model $\alpha$-attractor inflation model \cite{kalloshUniversalityClassConformal2013, turnerCoherentScalarfieldOscillations1983}. Since the black holes act as non-relativistic matter at large scales, if they are sufficiently long-lived, they will eventually dominate the Universe and lead to PBH domination. Hawking radiation converts the energy in black hole into radiation and completes the reheating \cite{hawkingParticleCreationBlack1975}. In this work, we expand on this simple picture by considering an additional accretion effect proposed in Ref.~\cite{dasImpactGeneralRelativistic2025}.

BHs could have formed during an epoch with the equation of state parameter $\omega$ where the perturbation of size of the horizon collapses shortly after the perturbation enters the horizon \cite{escrivaPrimordialBlackHoles2022, carrConstraintsPrimordialBlack2021}. Here only a monochromatic mass spectrum at formation is considered. The (initial) formation mass $m_f$ is a fraction of the mass contained in the horizon $m_H$ \cite{escrivaPrimordialBlackHoles2022}:
\begin{equation}
    m_f = \gamma m_H \equiv \gamma \frac{4\pi \mpl^2}{H_f}.
    \label{eq:Mf}
\end{equation}
The correction factor $\gamma$ accounts for the collapse efficiency. We take $\gamma \approx 0.2$ as reference value; in reality it depends on the equation of state, the amplitude and the shape of primordial perturbation \cite{escrivaPrimordialBlackHoles2022}. The abundance of BH at formation also needs to be specified; it is defined as the fraction of energy density at formation
\begin{equation}
    \beta \equiv \frac{\rho_{\text{PBH},f}}{3\mpl^2H_f^2}.
    \label{eq:beta}
\end{equation}
In this work, we remain agnostic about the formation scenario. Therefore, $m_f$ and $\beta$ are taken as free parameters.

We use the FLRW metric 
\begin{equation}
    \dd{s}^2 = \dd{t}^2 - a(t)^2 \dd{\vec{x}}^2 = a(\eta)^2 \left( \dd{\eta}^2 - \dd{\vec{x}}^2 \right),
\end{equation}
where the conformal time $\eta$ is defined. We use a dot to denote the derivative with respect to the cosmic time $t$. The Hubble parameter is defined as $H = \dot{a}/a$ and often the conformal Hubble parameter is used as well, $\mathcal{H} = aH = \dot{a}$. The set of equations governing the dynamics contains the equations for energy density in PBH, radiation and $\omega$-fluid, an equation for the evolution of the PBH mass, and finally the Friedmann equation:
\begin{align}
    \begin{split}
    \dot{\rho}_\text{PBH}+3H\rho_\text{PBH} &=  (-\Gamma_\text{evap} + \Gamma_\text{accret})\rho_\text{PBH}, \\
    \dot{\rho}_\text{rad}+4H\rho_\text{rad} &= \Gamma_\text{evap}\rho_\text{PBH}, \\
    \dot{\rho}_{\omega} + 3(1+\omega) H \rho_\omega  &= - \Gamma_{\text{accret}} \rho_\text{PBH}, \\
    \dot{m} &= -m\Gamma_\text{evap} + m\Gamma_\text{accret},  \\
    % = -A \frac{\mpl^4}{m^2} + \frac{\lambda_c \rho_\infty}{16\pi \mpl^4}m^2 , \\
    3\mpl^2H^2  &= \rho_\text{PBH} + \rho_\text{rad} + \rho_\omega .
    \end{split}
    \label{eq:backgrounds}
\end{align}
Note that $\Gamma_\text{accret}$ is omitted in the equation for $\rho_\text{rad}$, since $\rho_\text{rad}$ only comes from BH evaporation and it is negligible at early times. Here we use $\Gamma_\text{evap}$ to account for the semi-classical Hawking evaporation \cite{hooperDarkRadiationSuperheavy2019}
\begin{equation}
    \Gamma_\text{evap} = g_H \frac{\mathcal{G} \pi}{480} \frac{\mpl^4}{m^3} \equiv A \frac{\mpl^4}{m^3},
    \label{eq:Gamma-evap}
\end{equation}
with the greybody factor $\mathcal{G} = 3.8$. Only static Schwarzschild BHs are considered. $g_H$ counts the degrees of freedom below the black hole temperature $T_\text{BH}(m) = \mpl^2 / m$. For the black holes of interest ($M_\text{BH} \ll \SI{1e11}{\gram} $), it is constant $g_H \approx 108$\footnote{Not to be confused with the relativistic degrees of freedom in the thermal plasma: $g_*$ or $g_{s, *}$.}. The evaporation is considered complete at (assumes $\Gamma_\text{accret}=0$)~\cite{haquePrimordialBlackHole2023, domenechFormationEvaporationInduced2025}
\begin{equation}
    \teva(m) \approx \frac{m^3}{3 A \mpl^4}.
    \label{eq:eva-time}
\end{equation}
A competing effect with the Hawking radiation is accretion, described by the rate \cite{bondiSphericallySymmetricalAccretion1952, edgarReviewBondiHoyleLyttletonAccretion2004, kalitaRevisitingPBHAccretion2025}
\begin{equation}
    \Gamma_\text{accret} = \frac{\lambda_c \rho_\infty}{16\pi \mpl^4} m .
    \label{eq:Gamma-accret}
\end{equation}
Just like the classic Bondi-Hoyle accretion, the rate is proportional to the background density or the energy density at spatial infinity with respect to the BH~\cite{edgarReviewBondiHoyleLyttletonAccretion2004, bondiSphericallySymmetricalAccretion1952}; we take $\rho_\infty \rightarrow \rho_\omega$. Thus, accretion is most important shortly after formation when $\rho_\omega$ is maximal. The accretion coefficient $\lambda_c$ in Ref.~\cite{kalitaRevisitingPBHAccretion2025} is computed as a function of $\omega$
\begin{equation}
    \lambda_c = \frac{\omega^{-3/2}}{4}  (1+3\omega)^{(1+3\omega)/2\omega}.
    \label{eq:lambda-def}
\end{equation}
The final mass after accretion is given by (assume $\Gamma_\text{evap}=0$)~\cite{dasImpactGeneralRelativistic2025}
\begin{equation}
    m_\text{accret} \simeq m_f \left( 1 - \frac{\lambda_c \omega^{3/2}}{2(1+\omega)} \right)^{-1}.
\label{eq:m_accret}
\end{equation}
In the case of radiation domination ($\omega=1/3$), $\lambda_c=6\sqrt{3}$ and the black hole mass approaches $\sim 4 m_f$ before the eventual evaporation. Note that the RHS of Eq.~\eqref{eq:m_accret} has a pole for $\omega\to 1$.  This accreted mass is computed using the $\omega$ fluid as the sole component of the cosmic fluid and neglect the impact of BHs on the evolution of the cosmic fluid. For stiffer $\omega$, PBH domination is achieved earlier, as will be shown below, and the accretion is effectively halted during PBH domination. This divergence is thus not present in the PBH reheating scenario.
% Hence, for very stiff era, the accretion could potentially take longer to complete than for the PBHs to dominate the Universe. The equation of state parameter $\omega$ in Eq.~\eqref{eq:lambda-def} doesn't change in BHD, as it refers to the equation of state of the surrounding fluid (on small scales). 
% \cw{Is this pole "physical"? I believe this is done assuming only $\omega$. PBHD could change the result.}

For Eqs.~\eqref{eq:backgrounds} to be valid, we also need to make sure that the accretion radius is not larger than the cosmic horizon
\begin{equation}
    r_\text{accret} \sim \frac{m_f}{4\pi \mpl^2 c_s^2} \lesssim H_f^{-1} = \frac{m_f}{4\pi \gamma \mpl^2}.
\end{equation}
Additionally, we note that formation of PBH is not instant: the apparent horizon takes time to form \cite{escrivaPrimordialBlackHoles2022}. In such case, the cosmic horizon grows by a significant amount at formation time already. Taking the accretion effects at ``formation time'' into account can be thought as a simplification. On the other hand, the average separation of PBHs at formation is 
\begin{equation}
    d \equiv \lrfrac{3m_f}{4\pi \rho_{\text{PBH}, f}}^{1/3} = \frac{m_f}{4\pi \mpl^2 } (\gamma^2 \beta)^{-1/3},
\end{equation}
which is generally (much) greater than the accretion radius. Hence, accretion of PBH can be treated locally and isolated from other PBHs. This also justifies $\rho_\infty \rightarrow \rho_\omega$: the background fluid is largely unperturbed by accretion effect at large scales.

In this work, we focus on the possibility that the reheating is achieved by the thermal radiation released by the PBH via Hawking radiation. By equating the BH and radiation energy densities at evaporation time
\begin{equation}\label{eq:Trh-condition}
    {\rho}_\text{PBH} (\teva) \simeq \frac{3\mpl^2}{2}  H^2(\teva) \stackrel{!}{=} \rho_\text{rad} (\Trh) = \frac{\pi^2 g_* (\Trh)}{30} \Trh^4,
\end{equation}
we obtain an estimate for the reheating temperature
\begin{equation}
    \Trh \simeq \SI{0.6}{\mega \eV} \lrfrac{10^{9}\si{\gram}}{m}^{3/2} \lrfrac{106.75}{g_*(\Trh)}^{1/4} ,
    \label{eq:Trh}
\end{equation}
where we have used the maximal relativistic degrees of freedom in the Standard Model $g_*=106.75$ as the reference value. The BH mass $m$ here can be $m_f$ or $m_\text{accret}$ depending on whether accretion is included. The exact value of $\Trh$ can be computed from the solution to Eqs.~\eqref{eq:backgrounds} while taking the full temperature dependence in $g_*(T)$ into account.

From the Planck observation, we have an upper bound on the inflationary scale $H_\text{inf} < \num{2.5e-5}\mpl$ \cite{planck10}. This gives us a lower bound on the PBH mass, if they are indeed formed from the collapse of inflationary fluctuation. The evaporation timescale in Eq.~\eqref{eq:eva-time} can give the temperature at the onset of Hot Big Bang Universe. It must exceed $T_\text{BBN} \approx \SI{4}{\mega\eV}$ for BBN to proceed as in accordance with observation~\cite{kawasakiMeVscaleReheatingTemperature2000, hannestadWhatLowestPossible2004}. Together we have the rough estimates \footnote{As noted before, for $\omega \to 1$, the accretion mass in Eq.~\eqref{eq:m_accret} has to be taken with a grain of salt.}
\begin{equation}
    \SI{0.44}{\g} \lesssim m_f \lesssim \SI{3.6e8}{\g} \left( 1 - \frac{\lambda_c \omega^{3/2}}{2(1+\omega)} \right)^{3/2} .
\end{equation}
The PBH reheating temperature $\Trh \propto m_\text{accret}^{-3/2}$ contains the accreted mass and thus, the upper bounds gets reduced. This factor is generally $\sim \order{0.1}$ for $\omega \gtrsim 1/3$. This upper bound will be computed numerically from solving the background via Eq.~\eqref{eq:backgrounds} and Eq.~\eqref{eq:Trh-condition}. The upper bound on the PBH mass should be understood as requirement for a successful PBH reheating. Without the requirement of PBH reheating, the mere existence of PBH prior to BBN is less constrained (small abundance $\beta \lesssim \num{5e-20}$ is allowed \cite{wuPrimordialBlackHoles2025, carrConstraintsPrimordialBlack2021}).

\begin{figure}[ht]
    \centering
    \includegraphics[width=0.6\textwidth]{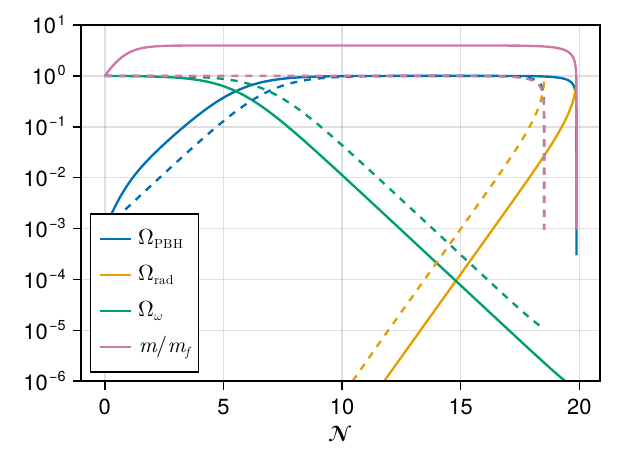}
    \caption{Background quantities from Eq.~\eqref{eq:backgrounds} for $\omega=1/3$ and $\beta=10^{-3}$. Horizontal axis shows the number of $e$-folds since BH formation $\mathcal{N}=\log(a/a_f)$. Blue, yellow, and green line correspond to the energy parameter of PBH, radiation from Hawking evaporation, and background fluid $\omega$. The purple line represents the mass of BH compared to the formation mass. Solid line is computed with accretion and $m_f=\SI{10}{\gram}$ and dashed line is without accretion and with $m_f=\SI{40}{\gram}$.}
    \label{fig:background}
\end{figure}
Now we want to solve the background equations~\eqref{eq:backgrounds} and check the effects of accretion. The naive expectation is that the accretion would only make the initial BH mass heavier, and it is just a simple matter of re-adjusting the initial BH mass for the final constraint on such scenario. We show an example of the solutions to the Eqs.~\eqref{eq:backgrounds} in Fig.~\ref{fig:background} with $\beta=10^{-3}$ and $\omega=1/3$. The number of $e$-folds since formation of BHs is used instead of the cosmic time
\begin{equation}
    \mathcal{N} \equiv \log \lrfrac{a}{a_f}.
\end{equation}
Here, we define the energy parameter of one component of the cosmic fluid as the fraction of the critical energy density
\begin{equation}
    \Omega_i \equiv \frac{\rho_i}{\rho_c} \equiv \frac{\rho_i}{3\mpl^2 H^2}.
\end{equation}
The different curves show various energy parameters: $\Omega_\text{PBH}$ (blue), $\Omega_\text{rad}$ (yellow), $\Omega_{\omega}$ (green), and finally $m/m_f$ (purple). Two scenarios are shown here: solid line is computed with accretion and $m_f=\SI{10}{\gram}$ and the dashed line without accretion but with the initial BH mass matching the accreted mass $m_f=\SI{40}{\gram}$. The general trends of two scenarios are similar: PBH gradually dominates the energy density after formation, while the background fluid ($\Omega_\omega$) redshifts away. At some point, the Hawking evaporation begins to be important: the radiation released by the evaporation starts to grow and finally is the only major source of energy density after the BH evaporates away and $m/m_f \to 0$. However, these two scenarios are quantitatively different. In the scenario with accretion (solid line), BH domination occurs earlier. We refer to the point where $\Omega_{\text{PBH}}(\teq) = \Omega_{\omega} (\teq)$ as the ($\omega$-BH-)equality. This is \textit{not} caused by the background fluid losing energy to BHs primarily, as in the first few $e$-folds $\Omega_\omega \sim 1$ still. Instead, the slower than $a^{-3}$ decrease of $\rho_\text{BH}$ due to accretion leads to the advancement of equality. One can also see that the evaporation time is delayed, even though they have the same effective mass. It can be understood as the effective formation time is delayed by the accretion effects. The net effect is then prolonging of the PBH domination phase. 

% We see that roughly an order of magnitude change to $\beta$ is necessary to compensate these effects. As shown by Ref.~\cite{domenechFormationEvaporationInduced2025}, the duration of the BH domination is crucial in determining the isocurvature induced gravitational waves upon evaporation.

This simple example shows that the accretion effects cannot be compensated simply by shifting the formation mass to accreted mass $m_f\to m_\text{accret}$. The other parameter, $\beta$, needs to be adjusted as well. As we will show below, $\beta$ needs to be shifted by $\order{10}$ to mimic the accretion effect.

\subsection{Analytical expression for the $\mathcal{N}_i$}\label{sec:back-ana}
Eqs.~\eqref{eq:backgrounds} can be solved numerically and one obtains the duration of each epoch. To have a better understanding, we attempt to give analytical expressions for the quantities. 

\begin{figure}[ht]
    \centering
    \includegraphics[width=0.5\textwidth]{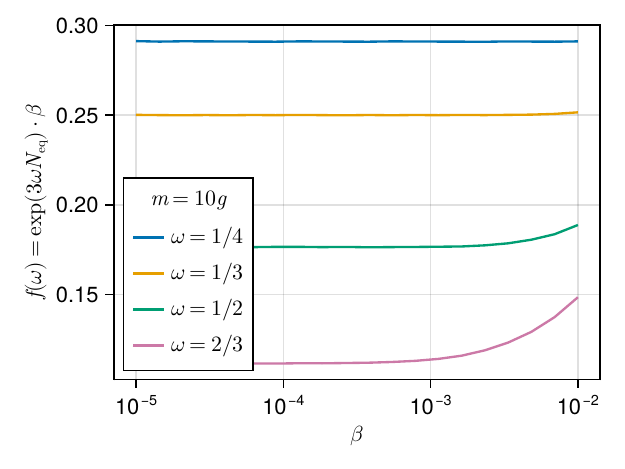}
    \caption{The fit function $\exp(3\omega \mathcal{N}_\text{eq})\beta = f(\omega)$ against $\beta$ for various values of $\omega$. Except for large $\omega$ and large $\beta$, $f=f(\omega)$ is only a function of $\omega$.}
    \label{fig:N_eq}
\end{figure}

Neglecting the effects of accretion and evaporation, $\mathcal{N}_\text{eq}$ can be easily obtained: $\mathcal{N}_\text{eq} \simeq -\log(\beta)/3\omega$ \cite{domenechGravitationalWaveConstraints2021}. Due to accretion, the duration of $\omega$-domination shortens, and we find the following estimate
\begin{equation}
    \mathcal{N}_\text{eq} = \frac{\log(f(\omega)/\beta)}{3\omega},
    \label{eq:N_eq}
\end{equation}
with the fitted function $f(\omega)\approx - 0.43\omega + 0.40$ as long as $\beta \lesssim 10^{-3}$. The duration of $\omega$-domination from numerical simulation of Eqs.~\eqref{eq:backgrounds} is shown in Fig.~\ref{fig:N_eq} for four values of $\omega$. It shows that the function $f$ decreases (roughly) linearly with $\omega$, for small $\beta$. This function $f(\omega)$ is unity without accretion, and it reflects the fact that accretion advances the $\omega$-BH-equality. As we will see, especially for large $\omega$, we are more interested in small $\beta \ll 1$. So this fit suffices for deriving the constraints on PBH reheating.

\begin{figure}[ht]
    \centering
    \includegraphics[width=0.9\textwidth]{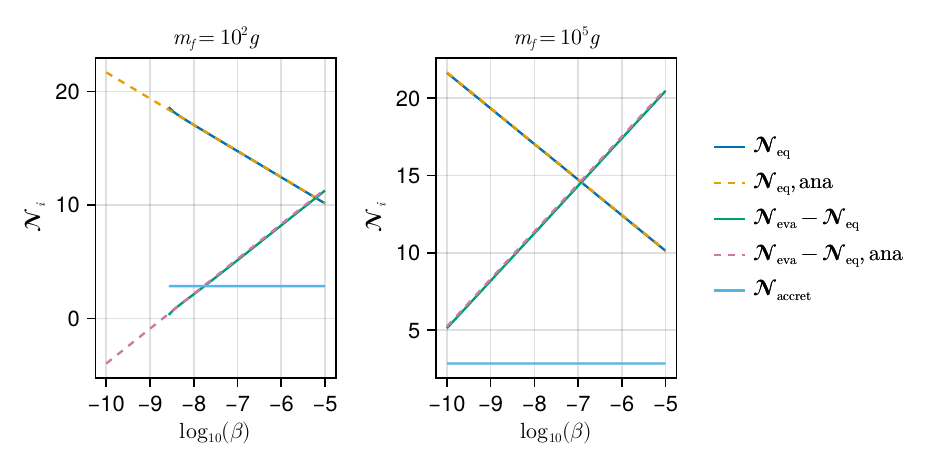}
    \caption{Duration of $\omega$-domination $\mathcal{N}_\text{eq}$ (blue), PBH domination $\mathcal{N}_\text{eva} - \mathcal{N}_\text{eq}$ (green) and completion of accretion $\mathcal{N}_\text{accret}$ (light blue) as function of $\beta$ with $m_f=10^2 \si{\g}$ (left) and $m_f=10^5 \si{\g}$. The equation of state parameter is fixed to $\omega=1/3$. Dashed yellow line shows the analytical fit Eq.~\eqref{eq:N_eq}. Dashed pink line shows Eq.~\eqref{eq:N_eva} with $m_f \to m_\text{accret}$ and $\beta \to 10\beta$.}
    \label{fig:Ns}
\end{figure}

Another important quantity is the evaporation time since the equality: $\mathcal{N}_\text{eva} - \mathcal{N}_\text{eq}$. In the scenario without accretion, it can be approximated as\footnote{$\mathcal{N}_\text{eq} = - \log(\beta)/3\omega$ is used and $\beta \ll 1$ is assumed. The quantity is rewritten first into $\mathcal{N}_\text{eva} - \mathcal{N}_\text{eq} = 2/3 \log(t_\text{eva}/t_\text{eq})$.} \cite{domenechFormationEvaporationInduced2025} 
\begin{equation}
    \mathcal{N}_\text{eva} - \mathcal{N}_\text{eq} \approx \frac{2}{3} \log( (1+\omega) \frac{2\pi\gamma}{A} \lrfrac{m}{\mpl}^2 \beta^{\frac{1+\omega}{2\omega}} ).
    \label{eq:N_eva}
\end{equation}
Accretion effect has two effects: BH-domination starting earlier and an ``effective'' delay of formation time. In practice, we find that rewriting $m_f \to m_\text{accret}$ and $\beta \to 10\beta$ gives a good description of numerical results. Fig.~\ref{fig:Ns} shows the quantities and their analytical fits. The dashed pink line and green solid line are this analytical description and numerical values for $\mathcal{N}_\text{eva} - \mathcal{N}_\text{eq}$ respectively. The yellow dashed line and blue solid line are the analytical and numerical values for $\mathcal{N}_\text{eq}$ with Eq.~\eqref{eq:N_eq}. These two analytical formulas give very good fit to the result. On the left, the numerical value stops at $\beta \gtrsim 10^{-9}$: beyond this value, there is no PBH domination. We also see that in the cases where the PBH domination barely happens, the accretion could take longer than equality time to complete (light blue curve)\footnote{$\mathcal{N}_\text{accret}$ is defined as the number of $e$-folds to achieve the maximal mass.}.

For PBH domination to happen, we need $\mathcal{N}_\text{eva} - \mathcal{N}_\text{eq} > 0$, which implies
\begin{equation}
    \beta \gtrsim \frac{1}{10} \left( \frac{1}{1+\omega} \frac{A}{2\pi \gamma} \lrfrac{\mpl}{m_\text{accret}}^2 \right)^{\frac{2\omega}{1+\omega}},
\end{equation}
where the constant $A$ has been defined in Eq.~\eqref{eq:Gamma-evap}. This minimal required $\beta$ generally gets reduced with the inclusion of accretion. 

Having computed the background evolution and found good analytical fits, we are going to consider observational consequences of accretion. 

\section{Isocurvature Induced Gravitational Waves}\label{sec:iso-gw}
We have solved the background quantities, and now it is time to investigate the observational constraints of PBH domination. Because of the sudden evaporation of the PBHs and termination of such early matter domination, a large isocurvature induced gravitational wave background could be produced \cite{domenechGravitationalWaveConstraints2021, papanikolaouGravitationalWavesUniverse2021}. This becomes an important constraint for PBH scenario.

Through the correlation function in Fourier space, we define $P$ as dimensionful power spectrum and $\mathcal{P}$ as the reduced (dimensionless) power spectrum
\begin{equation}
    \expval{X_{\vec{k}} X^*_{\vec{k}}} = P_X(k) \delta^{(3)} (\vec{k} - \vec{k}') = \frac{2\pi^2}{k^3} \mathcal{P}_{X}(k) \delta^{(3)} (\vec{k} - \vec{k}') .
\end{equation}
The PBHs are randomly formed in the Universe: finding a PBH at $\vec{x}$ is uncorrelated to finding another at $\vec{x}'$. At large scales, one can effectively treat the PBHs as a form of perfect fluid. The dimensionless power spectrum of PBH density $\delta \equiv \delta \rho_{\text{PBH}} / \rho_\text{PBH}$ is given by \cite{papanikolaouNewProbeNonGaussianities2024, papanikolaouGravitationalWavesUniverse2021, holstClusteringRunawayMerging2024}
\begin{equation}
    \mathcal{P}_\delta(k) = \frac{2}{3\pi} \lrfrac{k}{k_\text{uv}}^3 \Theta(k_\text{uv} - k) .
    \label{eq:P-delta-BH}
\end{equation}
The Heaviside function $\Theta$ makes the upper limit of the power spectrum explicit: the random distribution of PBH is only valid up to the average distance of the PBH. The UV scale is the comoving scale associated with the average distance between PBHs~\cite{papanikolaouNewProbeNonGaussianities2024}
\begin{equation}
    k_\text{uv} \equiv \frac{a_f}{d(a_f)} \equiv a_f \lrfrac{3 m_f}{4\pi \rho_{\text{PBH}, f}}^{-1/3} = a_f H_f \lrfrac{\beta}{\gamma}^{1/3},
\end{equation}
where Eqs.~\eqref{eq:Mf} and \eqref{eq:beta} have been used for the last equality.

We define the comoving scale which crossed the horizon at scale factor $a$ as $k=aH$. For later convenience, we show the various comoving scales necessary for the GW computation \cite{domenechFormationEvaporationInduced2025}. We obtain the following ratios of the scales
\begin{equation}
    \frac{k_\text{uv}}{k_f} = \lrfrac{\beta}{\gamma}^{1/3}, \quad
    \frac{k_\text{eq}}{k_f} \approx \sqrt{2} \lrfrac{\beta}{f(\omega)}^{\frac{1+3\omega}{6\omega}}, \quad
    \frac{k_\text{eva}}{k_f} \approx \left( \frac{5 \beta A}{\pi \gamma} \frac{\mpl^2}{m_\text{accret}^2} \right)^{1/3}.
    \label{eq:ks}
\end{equation}
For the second ratio, we have used Eq.~\eqref{eq:N_eq} and $H = 2/(3(1+\omega))t$.  The $\sqrt{2}$ comes from the fact that at $a=a_\text{eq}$, the PBHs have exactly the same energy density as the $\omega$-fluid: $\rho_\text{PBH}(a_\text{eq}) = \rho_\omega (a_\text{eq})$. For the last ratio, Eq.~\eqref{eq:N_eva} has been used and $(m, \beta)$ has been shifted to describe the effects of accretion.

Using the conformal Newtonian gauge, the general perturbed metric containing tensor and scalar perturbations can be written in conformal time $\eta$ as \cite{gorbunovIntroductionTheoryEarly2011} 
\begin{equation}
    \dd{s}^2 = a^2(\eta) \left[ (1+2\Psi) \dd{\eta}^2 - (\delta_{ij} + 2 \Phi \delta_{ij} + h_{ij}) \dd{x^i} \dd{x^j} \right].
    \label{eq:metric}
\end{equation}
where two scalar perturbations are related $\Psi = -\Phi$ in the absence of anisotropic stress. Here $h_{ij}$ describes the tensor degrees of freedom. The isocurvature perturbation describes the different composition of the cosmic fluid and is defined in our case as \cite{gorbunovIntroductionTheoryEarly2011, domenechGravitationalWaveConstraints2021}
\begin{equation}
    S \equiv \frac{\delta \rho_\text{PBH}}{\rho_\text{PBH}} - \frac{\delta \rho_\omega}{(1+\omega)\rho_\omega}.
\end{equation}
From the perturbations contained in the metric in Eq.~\eqref{eq:metric}, the $ij$-components of the perturbed Einstein field equation in Fourier space are found to be \cite{domenechFormationEvaporationInduced2025}
\begin{align}
    \Phi'' + 3 \mathcal{H}(1 + c_s^2) \Phi' + \left[ (1+3c_s^2)\mathcal{H}^2 + 2 \mathcal{H}' + c_s^2 k^2 \right] \Phi &= \frac{a^2 c_s^2}{2} \rho_\text{PBH} S,  \label{eq:Phi-diff-eq}\\
    S'' + (1+3(c_s^2 - \omega)) \mathcal{H}S' - k^2 (c_s^2 - \omega)S &= \frac{2c_s^2 k^4 }{a^2 (1+\omega) \rho_\omega} \Phi. \label{eq:S-diff-eq}
\end{align}
Derivative with respect to the conformal time is denoted with prime ${}'$. 
The (total) sound speed is given by 
\begin{equation}
    c_s^2  = \frac{c_\omega^2 (1+\omega)\rho_\omega}{\rho_\text{PBH}+(1+\omega)\rho_\omega}.
\end{equation}
For an ideal adiabatic fluid, the sound speed of the $\omega$-component equals the equation of state parameters $c_\omega^2 = \omega$ . 

There are two kinds of initial conditions: adiabatic with $\Phi_i \neq 0, S_i = 0$ and isocurvature $\Phi_i = 0, S_i \neq 0$ initial conditions. These two are linearly independent. We focus on the GWs from isocurvature initial conditions in this work. Studying the GWs from adiabatic initial conditions requires additional information on inflation, in particular the primordial curvature power spectrum $\mathcal{P}_\Phi$ at the corresponding scales. For pure isocurvature initial conditions, we find: $S \simeq \delta \rho_\text{PBH} / \rho_\text{PBH}$. Eqs.~\eqref{eq:Phi-diff-eq} and \eqref{eq:S-diff-eq} govern the evolution of the scalar perturbation and isocurvature perturbation and they are coupled through the right-hand sides of the equations. Hence, scalar perturbations can be generated via isocurvature perturbations (and vice versa). 

On super-(sound)horizon scales ($k \ll \mathcal{H}$), the isocurvature perturbation $S$ remains constant. Eq.~\eqref{eq:S-diff-eq} becomes $S'' + (1+3(c_s^2 - \omega)) \mathcal{H}S' = 0$, whose non-decaying solution is constant in time. On the other hand, the scalar perturbation remains constant on \textit{all scales} during matter domination \cite{gorbunovIntroductionTheoryEarly2011}, see Eq.~\eqref{eq:Phi-diff-eq} with $c_s=0$. Its value can be determined from the isocurvature perturbation $\Phi_\text{MD} \propto S$\cite{kodamaEvolutionIsocurvaturePerturbations1986}. After the sudden evaporation of BHs (most of the mass evaporate within $\order{1}$ e-folds at most, see Fig.~\ref{fig:background}) and the Universe enters RD, the gravitational potential $\Phi$ is still present, but there is non-negligible radiation pressure in Eq.~\eqref{eq:Phi-diff-eq}. The scalar perturbation begins to oscillate \cite{inomataEnhancementGravitationalWaves2019, domenechFormationEvaporationInduced2025}, similar to the baryonic acoustic oscillation around recombination \cite{gorbunovIntroductionTheoryEarly2011}. This sudden transition from BHD to RD enhances the amplitude of $\Phi$ \cite{domenechGravitationalWaveConstraints2021}, and the oscillation in $\Phi$ subsequently source the tensor perturbations. 

% Einstein field equation with metric \eqref{eq:metric} also leads to the equation of motion for tensor perturbations 
% \begin{equation}
%     h_k'' + 2 \mathcal{H}h_k' + k^2 h_k = 4 \mathcal{S}_k.
% \end{equation}
% with the tensor mode function $h$ and source term $\mathcal{S}$. \todo{Both needs to be defined} 
At linear order, perturbations of different helicity are decoupled. However, the tensor perturbations can be sourced at second order by scalar perturbation via the perturbed Einstein field equation. Schematically, the tensor power spectrum can be computed as~\cite{kohriSemianalyticCalculationGravitational2018}
\begin{equation}
    \mathcal{P}_h \sim \int \dd{k} \int \dd{k}' \left( \int \dd{t}f(k, k', t) \right)^2 \mathcal{P}_\Phi(k) \mathcal{P}_\Phi(k'),
    \label{eq:P_h}
\end{equation}
where $f(k, k', t)$ is some oscillating function, which is in turn an integral of the Green's function and source term in the differential equation for $h_{ij}$ \cite{domenechGravitationalWaveConstraints2021}. The energy density parameter of the GW can be obtained from the oscillation average of $\mathcal{P}_h$ \cite{kohriSemianalyticCalculationGravitational2018, domenechFormationEvaporationInduced2025}
\begin{equation}
    \Omega_\text{GW}(k, \eta) = \frac{k^2}{12 \mathcal{H}^2} \overline{\mathcal{P}_h(k, \eta)}.
    \label{eq:Omega_gw}
\end{equation}

Ref.~\cite{domenechFormationEvaporationInduced2025} has solved Eqs.~\eqref{eq:Phi-diff-eq} and \eqref{eq:S-diff-eq}, and found good analytical fits using pure isocurvature initial conditions. Subsequently, the power spectrum and energy parameter of the tensor perturbations are derived using Eqs.~\eqref{eq:P_h} and \eqref{eq:Omega_gw}. The resulting GW spectrum has a characteristic power law behaviour $\Omega \propto k^{11/3}$ up to the UV scale $k_\text{uv}$. The overall amplitude depends on the various scales shown in Eq.~\eqref{eq:ks}. One finds a simple analytic expression for the energy parameter at BBN \cite{domenechFormationEvaporationInduced2025}
\begin{equation}
    \Omega_\text{GW}(k) \approx \Omega_{\text{GW, peak}} \lrfrac{k}{k_\text{uv}}^{11/3} \Theta(1-k/k_\text{uv}),
\end{equation}
where the peak value can be written as 
\begin{align}
    \Omega_\text{GW, peak} &\approx C(\omega)^4 \frac{c_s^{7/3}(c_s^2 - 1)^2}{576 \cdot 6^{1/3} \pi} \lrfrac{k_\text{eq}}{k_\text{uv}}^8 \lrfrac{k_\text{uv}}{k_\text{eva}}^{17/3}  , \label{eq:Omega_iso} \\
                           &\approx \num{9.58e30} \cdot C^4(\omega) \left( 1 - \frac{\lambda_c \omega^{3/2}}{2(1+\omega)}\right)^{-34/9}  (10\beta)^{\frac{4(1+\omega)}{3\omega}} \notag \\
                           &\quad \times \lrfrac{\gamma}{0.2}^{8/3} \lrfrac{m_f}{10^4 \si{\gram}}^{34/9},
    % \Omega_\text{GW, BBN}^\text{tot} \approx 0.31 \cdot C(\omega)^4 \frac{c_s^{7/3}(c_s^2 - 1)^2}{576 \cdot 6^{1/3} \pi} \lrfrac{k_\text{eq}}{k_\text{uv}}^8 \lrfrac{k_\text{uv}}{k_\text{eva}}^{17/3}.
    \label{eq:Omega_iso_approx}
\end{align}
with the sound speed in radiation domination $c_s = 1/\sqrt{3}$. The function $C(\omega)$ is fitted to be
\begin{equation}
    C(\omega) = \frac{9}{20} \alpha^{-1/3\omega} \left( 3 + \frac{1-3\omega}{1+3\omega} \right)^{-1/3\omega},
\end{equation}
with $\alpha \approx 0.135$. The factor $k_\text{eq} / k_\text{uv}$ in Eq.~\eqref{eq:Omega_iso} reflects the effects of the initial PBH abundance and when the equality takes place relative to the formation time. The other factor $k_\text{uv} / k_\text{eva}$, although it doesn't depend on $\beta$ at all \cite{domenechGravitationalWaveConstraints2021}, correlates positively with the duration of PBH domination. This can be most easily understood in terms of the density contrast $\delta = \delta \rho / \rho$, which grows linearly with the scale factor in MD. In the end, $\Omega_\text{GW, peak}$ strongly positively correlates with $\beta$ and $m_f$. 

The GWs produced right after evaporation propagate as relativistic particles and their energy density redshifts in the same way as the background radiation. In order not to spoil the standard BBN predictions, there is an upper limit on the radiation component in addition to the Standard Model particles. The GW must satisfy at BBN \cite{capriniCosmologicalBackgroundsGravitational2018}
\begin{equation}
    \Omega_\text{GW}^\text{tot} \equiv \int \dd{\ln(k)} \Omega_\text{GW}(k) \lesssim \frac{7}{8} \lrfrac{4}{11}^{4/3} \Delta N_\text{eff} \approx 0.068.
    \label{eq:Omega_bbn}
\end{equation}
Once the background is computed via Eq.~\eqref{eq:backgrounds}, the scales in Eq.~\eqref{eq:ks} can be obtained. Then one can check if the BH reheating scenario with the given $(m_f, \beta)$ is compatible with the BBN prediction via Eqs.~\eqref{eq:Omega_iso} and \eqref{eq:Omega_bbn}.

\begin{figure}[ht]
    \centering
    \includegraphics[width=0.5\textwidth]{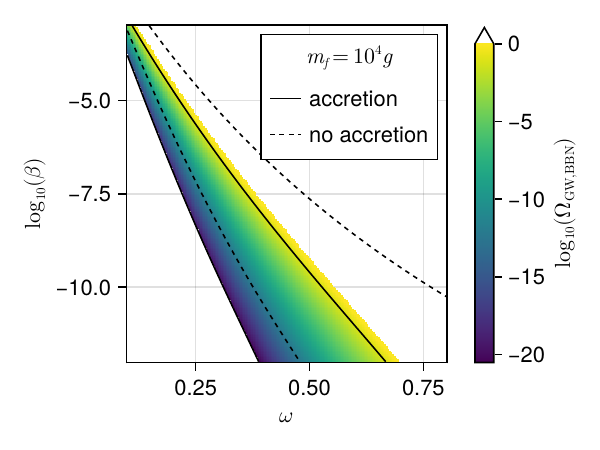}%
    \includegraphics[width=0.5\textwidth]{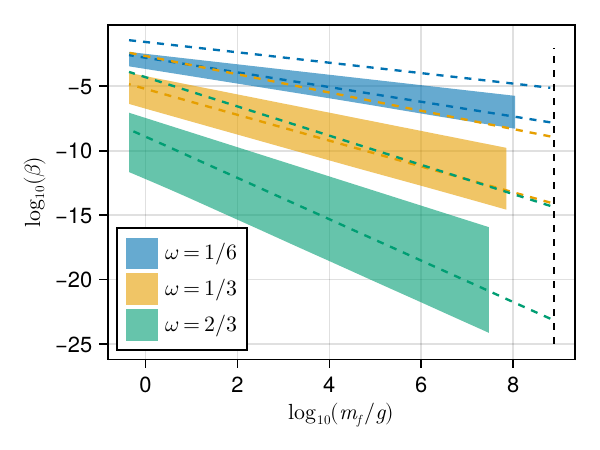}
    \caption{BBN constraints. Left: the GW energy density parameter $\Omega_\text{GW, BBN}$ in relation to $\beta$ and $\omega$ with $m_f=10^{4}\si{\gram}$. The region enclosed by lines corresponds to the situation where the GW is not overproduced and PBH-domination takes place. Solid (dashed) lines are for the scenarios with(out) accretion. Right: Allowed parameter region for PBH domination for various values of $\omega$. Regions enclosed by dashed lines are allowed regions without accretion.}
    \label{fig:BBN-cons}
\end{figure}

Fig.~\ref{fig:BBN-cons} shows the BBN constraint on isocurvature induced GW from PBH reheating. On the left, the GW energy density at BBN $\Omega_\text{GW, BBN}$ is plotted as function of two input parameters $\beta$ and $\omega$, with a single fixed $m_f=10^{4}\si{\gram}$. The lower bound in $\beta$ is only to ensure PBH reheating, while the upper bound corresponds to overproduction of $\Omega_\text{GW}$ leading to violate the BBN $\Delta N_\text{eff}$ bound. Compared to the treatment without accretion (shown with dashed lines), the bounds with accretion (solid lines) move to smaller $\omega$ and lower $\beta$. 
On the right, we have selected a few $\omega$'s and varied the formation PBH mass $m_f$. The allowed parameter space with PBH reheating is shown with shaded regions (with accretion) and regions enclosed by the dashed lines (without accretion). The lower bound on $m_f$ comes from the constraint on the inflationary scale: it is universal and unaffected by the accretion. The upper bound of $m_f$, however, comes from the condition of successful BBN, $\Trh \gtrsim \SI{4}{\mega \eV}$. This is also universal for all $\omega$ without accretion. Accretion wound lead to heavier BH and thus lower reheating temperature, depending on the specific $\omega$. The upper bounds on $m_{f}$ is reduced in such cases. As observed in Section \ref{sec:back-ana}, the effects of accretion also effectively increase $\beta$ by about an order of magnitude, in terms of the background evolution. Thus, the allowed regions are all roughly shifted downwards by a similar amount. Altogether, we see that the effects of accretion cannot be simply absorbed by adjusting the formation mass to the accretion mass $m_f \to m_\text{accret}$, as indicated by the background evolution already. The net effect of accretion is more powerful than one would naively expect.

The isocurvature induced GWs can be potentially detected by future experiments, as the frequency lies in the $\sim [1, 10^7] \, \si{\hertz}$ range~\cite{domenechFormationEvaporationInduced2025}:
\begin{equation}
    f_\text{uv} = \frac{k_\text{uv}}{2\pi a_0} \simeq \SI{7.3}{\kilo\hertz} \cdot  \tilde{f}(\omega) (1+\omega)^{-2/3} \lrfrac{3.91}{g_{*s}(\Trh)}^{1/3} \lrfrac{10^4\si{\gram}}{m_f}^{-5/6},
    \label{eq:f_UV}
\end{equation}
where entropy conservation is used for $a_\text{eva}/a_0$, and we have used $g_{*} \simeq g_{*, s}$. The relativistic degree of freedom $g_*(T)$ is taken from Ref.~\cite{husdalEffectiveDegreesFreedom2016}. The auxiliary function $\tilde{f}(\omega)$ is defined as 
\begin{equation}
    \tilde{f}(\omega) = 10^{-\frac{1+\omega}{3\omega}} (f(\omega))^{-\frac{1}{3\omega}}  \left( 1 - \frac{\lambda_c \omega^{3/2}}{2(1+\omega)} \right)^{\frac{17}{6}},
\end{equation}
to capture the effects of accretion. Note that $f_\text{uv}$ in Eq.~\eqref{eq:f_UV} depends on PBH formation abundance $\beta$ only through $\Trh$; it thus depends on $\beta$ very weakly.

The current day GW energy parameter can also be computed:
\begin{equation}
    \Omega_{\text{GW}, 0}(k) h^2 = 0.387 \lrfrac{g_*(\Trh)}{106.75}^{-1/3} \Omega_{r,0}h^2 \; \Omega_\text{GW}(k),
\end{equation}
with $\Omega_{r,0}h^2 = \num{4.18e-5}$ the current-day radiation energy parameter \cite{planck6}.

\begin{figure}[ht]
    \centering
    \includegraphics[width=0.7\textwidth]{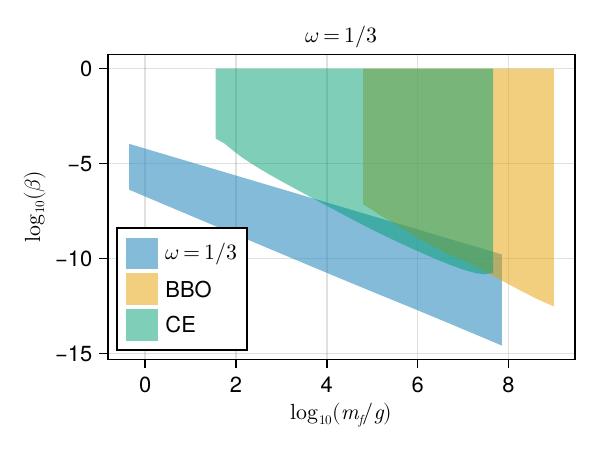}
    \caption{Allowed parameter regions with accretion and $\omega=1/3$ in blue. Green and yellow areas correspond to the parameter regions where the Cosmic Explorer (CE) and Big Bang Observatory (BBO) are able to detect the isocurvature induced GWs, respectively.}
    \label{fig:BBN-cons-1-3}
\end{figure}

In some cases, future detectors in their corresponding frequency ranges are foreseen to provide better sensitivities in $\Omega_\text{GW}$ than the BBN bound. Hence, such detectors could bring light on the PBH reheating scenario. One can check if certain signal $\Omega_\text{signal} = \Omega_{\text{GW}, 0}$ can be detected by the detector, characterized by the detector sensitivity curve $\Omega_\text{noise}$, by computing the signal-to-noise ratio \cite{schmitzNewSensitivityCurves2021, papanikolaouGravitationalWavesUniverse2021}
\begin{equation}
    \text{SNR}^2 = n_\text{det} t_\text{obs} \int \dd{f} \lrfrac{\Omega_\text{signal}(f)}{\Omega_\text{noise}(f)}^2,
    \label{eq:SNR}
\end{equation}
where $n_\text{det} = 1$($2$) for an auto-correlation (cross-correlation) type detector and $t_\text{obs}$ is the observation time. We take $t_\text{obs}= 1 \text{yr}$ as reference value\footnote{Since we expect $\Omega_\text{GW}$ to be generally exponential in $\beta$ and $m_f$, any realistic value of $t_\text{obs}$ would not change the (would-be) constraints significantly.}. 
The detector sensitivity curve $\Omega_\text{noise}$ is taken from Ref.~\cite{schmitzNewSensitivityCurves2021} and the signal-to-noise ratio is computed using Eq.~\eqref{eq:SNR}. Big-Bang Observer (BBO) and Cosmic Explorer (CE) are two most sensitive detectors in the frequency range of interest. Their sensitivity curves are projected to the $(m_f, \beta)$ plane in Fig.~\ref{fig:BBN-cons-1-3}. The upper right colored regions (green and orange) are the parameter region, which these two detectors can in principle probe. Both detectors are able to probe upper-right (large $m_f$ and $\beta$) corner of the allowed parameter space (light blue). CE is more sensitive at large frequency, hence it can probe the scenario with lower $m_f$. Note that the projected sensitivity curve relies on the underlying physics model. We find the two detectors can only probe the upper-right corner of the allowed parameter region shown in Fig.~\ref{fig:BBN-cons}, for all values of $\omega$ regardless of consideration of accretion.

Before moving on to the next section, there are a few comments. 
We have only considered GWs from pure isocurvature initial conditions. Additional GWs from adiabatic initial conditions are always present and introduce a lower frequency component \cite{domenechFormationEvaporationInduced2025, papanikolaouGravitationalWavesUniverse2021, inomataGravitationalWaveProduction2020}, but one requires additional information on the inflation model for precise prediction. 
In this Section, we have considered truly random or uncorrelated PBH distribution in Eq.~\eqref{eq:P-delta-BH}. This is no longer the case, if there are sizable non-gaussianities present. In such cases, the PBH number density would be correlated and one has to include an additional factor in Eq.~\eqref{eq:P-delta-BH}. The GWs would have a secondary peak at lower frequencies \cite{domenechGravitationalWaveConstraints2021, papanikolaouNewProbeNonGaussianities2024}. 
Furthermore, if the BHs are formed from inflationary inhomogeneities, the required primordial power spectrum would be around $\mathcal{P}_\zeta \sim 10^{-2}$ \cite{escrivaPrimordialBlackHoles2022}. Such fluctuations would inevitably also lead to scalar induced gravitational waves $\Omega_\text{SIGW} \simeq \mathcal{P}_\zeta^2$ at formation \cite{capriniCosmologicalBackgroundsGravitational2018}. These GWs, however, would be diluted by the early matter domination if the PBH domination indeed takes place and not relevant for the BBN constraints. The frequency of such GWs is even higher than the isocurvature induced GWs, see Eq.~\eqref{eq:ks}.
We also want to point out that in the usual accretion derivation, the background fluid is assumed to be at rest with respect to the BH (at spatial infinity) \cite{dasImpactGeneralRelativistic2025}. This relative velocity $V_\text{rel}$ can be in principle calculated from the isocurvature perturbation using $S' = k^2 V_\text{rel}$. The correction due to $V_\text{rel} \neq 0$ is only expected to be significant for relativistic relative velocity $V_\text{rel} \to 1$.

In addition to the isocurvature induced GWs, there are also gravitons emitted from Hawking radiation: they are suppressed by $\order{10^{-2}}$ compared to a scalar or fermion. This is a pure spin effect: the effective potential in the Schroedinger-like equation (Teukolsky equation) governing the field around the BH contains a spin term explicit which suppresses graviton emission~\cite{pageParticleEmissionRates1976}. So effectively they contain $\order{10^{-4}}$ of the energy density of the BHs at evaporation. Since gravitons redshifts just like the other particles behaving like radiation, this gives $\order{10^{-4}}$ contribution to $\Delta N_\text{eff}$. The gravitons from Hawking radiation indeed lead to a worse constraint for BH reheating scenario, compared to the isocurvature induced gravitational waves discussed in this Section.

In this Section, we have considered that the PBH perturbation induced GWs at large scales after the evaporation. We find accretion effects significant in shifting the allowed parameter region. Future experiments are expected to probe a fraction of such region.

\section{Merging of BHs in PBH domination}
\label{sec:merger}
The discussion in Section \ref{sec:iso-gw} is entirely based on linear perturbation theory. It is valid on large scales where the density contrast remains below unity throughout the BH domination. The density contrast grows linearly in matter domination $\delta = \delta \rho / \rho \propto a(t)$. As the PBH domination can last $\sim \order{10}$ $e$-folds, it is expected for the BHs to form clusters just like structure formation in the usual matter domination. The BHs in the clusters could then potentially merge, thereby producing heavier BHs. Here we consider a simplified description of such scenario with some analytical approximations based on Ref.~\cite{holstClusteringRunawayMerging2024}. Cluster formation is modeled using the Press-Schechter formalism. Each cluster is treated in isolation and there are two competing effects: merging and cluster evaporation. This simple description doesn't track the BH mass distribution within a cluster, and both effects only depend on the average BH mass. As can be seen from Fig.~\ref{fig:Ns}, BH accretion usually completes before the equality, unless the PBH barely dominate the Universe. As we shall see, BH merging is significant only for relatively large $\beta$. Thus, in this Section, we assume accretion always completes before BH domination. As already being noted in Section \ref{sec:background}, the inclusion of accretion introduces a non-trivial change to the background evolution. In this Section, we consider the effect of structure formation and merging on the final black hole mass spectrum in the presence of accretion. The final constraint on PBH reheating scenario comes from the extended BH mass function as a result of merging. We use $M$ exclusively to denote the BH cluster mass, to make it distinct from the mass $m$ of a single BH.

The basis of the merging computation is structure formation during early matter domination. Just like the galaxies we see today, the PBHs form clusters according to the usual Press-Schecter (PS) formalism. From the linear power spectrum, one can obtain the mass function of collapsed objects at a particular time\footnote{Note that we put the time-dependence in $\sigma$, instead of in $\delta_c$.}~\cite{moGalaxyFormationEvolution2010, schneiderExtragalacticAstronomyCosmology2015, binneyGalacticDynamics2008, coorayHaloModelsLarge2002}
\begin{equation}
    % \dv{n_{\text{cl}}(M, t)}{M} = \sqrt{\frac{2}{\pi}} \frac{\rho_\text{BH}(t)}{M^2} \frac{\delta_c}{\sigma (M, t)} \exp(-\frac{\delta_c^2}{2\sigma(M, t)^2}) \left| \dv{\ln \sigma(M,t)}{\ln M} \right|,
    \dv{n_{\text{cl}}(M, t)}{M} = \sqrt{\frac{2}{\pi}} \frac{\rho_\text{BH}(t)}{M^2} \nu \exp(-\nu^2/2) \left| \dv{\ln \nu}{\ln M} \right|,
    \label{eq:PS-mass-function}
\end{equation}
where $\rho_\text{BH}(t) = m_f n_\text{BH}(t)$ is the \textit{background} energy density of the black holes and $\nu\equiv \delta_c/\sigma(M, t)$. For spherical collapse, the critical density contrast $\delta_c \approx 1.69$. The mass variance of the smoothed density field characterized by the linear matter power spectrum is defined as 
\begin{equation}
    \sigma^2(M, t) = \int_0^{\infty} \mathcal{P}_\delta (k, t) \tilde{W}^2(kR) \dd{\ln k},
    \label{eq:sigma-def}
\end{equation}
where the BH density power spectrum in Eq.~\eqref{eq:P-delta-BH} grows in matter domination: 
\begin{equation}
\mathcal{P}_\delta (k, t) = 
\begin{cases}
    \mathcal{P}_\delta(k) (a/\aeq)^2 , & k > k_\text{eq},  \\
    \mathcal{P}_\delta(k) (a/a_k)^2 , & k \leq k_\text{eq},
\end{cases}    
\end{equation}
because fluctuations only start growing only after they cross the horizon ($k = a_k H(a_k)$).
$k$-space top hat window function is used here $\tilde{W}(x) = \Theta(1-x)$. The smoothing radius with such window function is related to the mean mass contained in it by \cite{moGalaxyFormationEvolution2010}
\begin{equation}
    R (M) = \lrfrac{M}{6\pi^2 \rho_{\text{BH}}}^{1/3}.
    \label{eq:smoothing-radius}
\end{equation}
The mass variance defined in Eq.~\eqref{eq:sigma-def} can be computed using the power spectrum in Eq.~\eqref{eq:P-delta-BH} with the growth factor $(a/a_\text{eq})^2$ \cite{holstClusteringRunawayMerging2024}:
\begin{equation}
    \sigma^2(M, t) = \frac{m_\text{accret}}{M} \lrfrac{a}{a_\text{eq}}^2 \mu(M),
    \label{eq:sigma-real}
\end{equation}
with
\begin{equation}
    \mu(M) = 
    \begin{cases}
        \frac{3}{7} \lrfrac{M}{\MHeq}^{-4/3}, & M > \MHeq, \\
        1-\frac{4}{7}\frac{M}{\MHeq}, & M < \MHeq.
    \end{cases}
    \label{eq:mu-definition}
\end{equation}
% It has been used that the density contrast at scales larger than the horizon size at equality ($k < \keq$) grow only $ \propto a/a_k$ instead of $ \propto a/\aeq$, thus the time-dependent power spectrum has $(k/\keq)^{4}$ suppression at scales larger than $\keq$. 
The case distinction arises because $\keq$ can be either larger or smaller than the cutoff imposed by the window function $R(M)^{-1}$. The horizon mass at equality $\MHeq$ is given via Eqs.~\eqref{eq:Mf}, \eqref{eq:beta}, and \eqref{eq:N_eq} as 
\begin{equation}
    \MHeq \simeq \frac{m_f}{\gamma} \lrfrac{f(\omega)}{\beta}^{\frac{1+\omega}{2\omega}}.
\end{equation}
With the mass function in Eq.~\eqref{eq:PS-mass-function}, the Press-Schecter formalism shows how clusters develop: a cluster with mass $M$ forms at $\sigma(M, t) \simeq \delta_c$ and small clusters form early. 

We approximate BH cluster of total mass $M$ by an isotropic singular isothermal sphere. The constituents follow a Maxwellian velocity distribution. The velocity dispersion of such distribution is \cite{cerdenoDirectDetectionWIMPs2010, binneyGalacticDynamics2008}
\begin{equation}
    % v \approx \frac{4}{\sqrt{\pi}} \sigma_v = \frac{1}{\pi\mpl} \sqrt{\frac{2 M }{R_\text{cl}}},
    \sigma_v = \frac{1}{\mpl} \sqrt{\frac{M}{8\pi R_\text{cl}}},
\end{equation}
where the cluster radius is taken as\footnote{Not to be confused with the smoothing radius in equation \eqref{eq:smoothing-radius}.}
\begin{equation}
    R_\text{cl} = \lrfrac{3 M}{4\pi \rho_\text{cl}}^{1/3}.
\end{equation}
According to the simple assumptions of spherical collapse and subsequent virialization, the cluster energy density is roughly $18 \pi^2$ times that of the background BH density $\rho_\text{BH}$~\cite{schneiderExtragalacticAstronomyCosmology2015, moGalaxyFormationEvolution2010}. Thus, the initial cluster energy density at cluster formation $a_\text{cl}/a_\text{eq} = \delta_c \sqrt{N_i / \mu(M_i)}$, where $N_i$ denotes the (initial) number of BHs in cluster at cluster formation, reads
\begin{equation}
    \rho_{\text{cl}}(a_\text{cl}) \simeq 18\pi^2 \rho_\text{BH}(a_\text{cl}) = 18\pi^2 \rho_\text{BH}(a_\text{eq}) \lrfrac{\mu(M_i)}{ \delta_c^2 N_i}^{3/2}.
    \label{eq:rho_cl}
\end{equation}
Here, the initial mass can be expressed as $M_i = M(a_\text{cl}) = N_i m_\text{accret} $. The background BH energy density at the equality can be obtained via Eq.~\eqref{eq:Mf} and Eq.~\eqref{eq:N_eq}:
\begin{equation}
    \rho_\text{BH}(\aeq) 
        % &= 3\mpl^2 H_f^2 \beta \lrfrac{\aeq}{a_\text{f}}^{-3} \notag \\
      % &= 3\mpl^2 H_f^2 \beta e^{-3 N_\text{eq}} \\
    \simeq 48 \pi^2 \gamma^2 \beta \frac{\mpl^6}{m_f^2} \lrfrac{f(\omega)}{\beta}^{-1/\omega} .
\end{equation}

When two BHs approach each other, there is a time-dependent quadrupole moment and energy is radiated away via GW. There exists a threshold impact parameter so that these two BHs lose enough energy and merge instead of having a close encounter. This two-body process can be modeled as particle scattering and the merging cross-section is \cite{mouriRunawayMergingBlack2002}
\begin{equation}
    \sigma_{2b} = 128\pi^3 \lrfrac{85\pi}{6\sqrt{2}}^{2/7} \frac{(m_i + m_j)^{10/7} m_i^{2/7} m_j^{2/7}}{\mpl^4 v^{18/7}},
\end{equation}
with $v = (4/\sqrt{\pi})\sigma_v$ the average velocity for the Maxwellian distribution \cite{binneyGalacticDynamics2008}. The merging rate within a cluster of mass $M$ and size $N$ is given by \cite{holstClusteringRunawayMerging2024}
\begin{equation}
    \Gamma_\text{merging}(M, N) = \frac{N}{2} n_{\text{cl}, \text{BH}}\sigma_{2b}(M/N, v) v.
\end{equation}
Here, $n_{\text{cl},\text{BH}} = \rho_\text{cl} / (M/N)$ is the average number density of BH in a cluster. 

After cluster formation, there is a probability that a member of the cluster escapes via ejection (a close encounter) or evaporation (a series of weak encounters). The latter is typically more dominant and will be considered here \cite{binneyGalacticDynamics2008}. At the tail of the Maxwellian velocity distribution, the BH could have enough velocity to escape the cluster and this evaporation rate is \cite{binneyGalacticDynamics2008, holstClusteringRunawayMerging2024}
\begin{equation}
    \GammaE (M, N) = 7.4 \cdot 10^{-3} \cdot \frac{N}{t_\text{relax}} = 7.4 \cdot 10^{-3} \cdot \frac{8 \ln(N) \sigma_v}{R}.
\end{equation}
After the clusters are formed, their energy density can change due to BHs escaping the cluster. It is assumed that there is mass loss but no total energy loss. Thus, the cluster energy density in Eq.~\eqref{eq:rho_cl} scales as $M^{-5}$ as the total mass after its formation \cite{holstClusteringRunawayMerging2024}.

If we treat the clusters as isolated, then the total cluster mass $M$ and total number of PBH in the cluster $N$ can be described by the following differential equations \cite{holstClusteringRunawayMerging2024}:
\begin{align}
    \begin{split}
    \dv{N}{t} &= -\GammaM -\GammaE, \\
    \dv{M}{t} &= - \Delta \frac{M}{N} \GammaM - \frac{M}{N} \GammaE - N \dot{m},
    \end{split}
    \label{eq:diff_eq_M_N_orig}
\end{align}
with $\Delta$ parametrizing the fraction of loss of PBH mass into GW radiation and is typically $\Delta \simeq 0.05 $~\cite{collaborationObservationGravitationalWaves2016}. The number of PBH in one cluster is decreased by merging events (characterized by $\GammaM$) and cluster evaporation (characterized by $\Gamma_\text{cl, evap}$). Total cluster mass is reduced via cluster evaporation and merging events, where part of the total mass is radiated away via GW. To simplify the problem, the distribution of PBH masses within the cluster is not tracked. The aforementioned processes are characterized by the mean PBH mass in the cluster $M / N$. 

To further simplify the equations, one can omit the Hawking radiation term in Eq.~\eqref{eq:diff_eq_M_N_orig} and terminate the integration at the evaporation time. We note that from Eq.~\eqref{eq:N_eq} the equality time can be written as 
\begin{equation}
    t_\text{eq} \simeq \frac{2}{3(1+\omega)} \frac{M_f}{4\pi\mpl^2 \gamma} \lrfrac{f(\omega)}{\beta}^{(1+\omega)/2\omega}.
    \label{eq:t_eq_clustering}
\end{equation}
Then the ratio of equality time and evaporation time is given by
\begin{equation}
    \frac{t_\text{eq}}{\teva} \simeq \frac{A}{2\pi \gamma (1+\omega)} \lrfrac{\mpl}{m_f}^2 \lrfrac{f(\omega)}{\beta}^{(1+\omega)/2\omega}.
\end{equation}
As we shall see, the lowest relevant $\beta$ for merging constraint is $\sim 10^{-10}$ in which case $t_\text{eq} \ll \teva$. Thus, the integration of Eq.~\eqref{eq:diff_eq_M_N_orig} can be safely terminated at $\teva$. For small clusters, it could be unnecessary to continue the integration once there is only one BH left.

The solutions of Eq.~\eqref{eq:diff_eq_M_N_orig} establish how the initial cluster size $N_i$ relates to the final relic BH mass $m_\text{relic}$. Eq.~\eqref{eq:diff_eq_M_N_orig} can be reformulated into a single differential equation with $N$ being the independent variable, if the Hawking evaporation is omitted and we assume that $M/N$ doesn't change dramatically. This leads to the analytical solution~\cite{holstClusteringRunawayMerging2024}
\begin{equation}
    m_\text{relic} = M_i \lrfrac{\Gamma_{\text{cl, evap, } i} + \Gamma_{\text{merging, }i} N_i^{5/7}}{\Gamma_{\text{cl, evap, } i} + \Gamma_{\text{merging, } i}}^{\frac{7}{5}(1-\Delta)},
    \label{eq:m_relic}
\end{equation}
where the quantities with subscript $i$ are the initial values (taken at cluster formation time). Both the full treatment using Eq.~\eqref{eq:diff_eq_M_N_orig} and the analytical approximation Eq.~\eqref{eq:m_relic} must have an upper limit for the relic PBH mass $N_\text{max}$, as the BH domination comes to an end due to the Hawking evaporation. Large clusters form late, hence their constituents could never have merged. This can be done by comparing the merging or cluster evaporation timescale with the BH evaporation timescale for the analytical treatment~\cite{holstClusteringRunawayMerging2024}. Numerically solving Eq.~\eqref{eq:diff_eq_M_N_orig} automatically incorporates such an upper limit.
% Thus, we need 
% \begin{equation}
%     \min(t_\text{merging}, t_\text{cl, evap}) < t_\text{eva}
% \end{equation}
% where 
% \begin{equation}
%     t_\text{merging} \simeq \frac{N_i}{\GammaM (N_i)}, \quad t_\text{cl, evap} \simeq \frac{2N_i}{7 \GammaE (N_i)}
% \end{equation}

\begin{figure}[ht]
    \centering
    \includegraphics[width=0.6\textwidth]{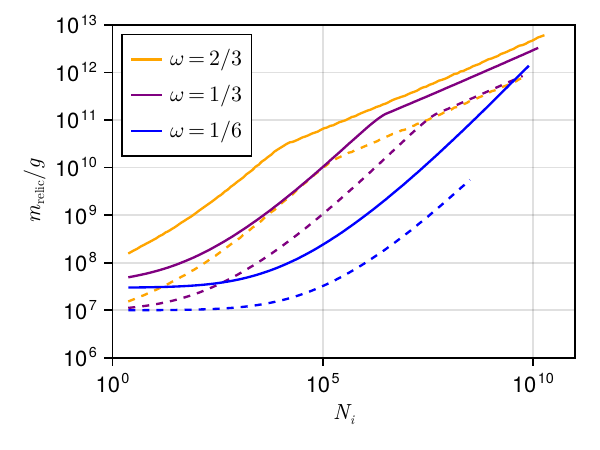}
    \caption{Relic BH mass $m_\text{relic}$ at BH evaporation time against initial cluster size $N_i$. Solid (dashed) lines are drawn with(out) accretion effects. Equation of state parameter $\omega$ prior to the BH domination is varied.}
    \label{fig:avrg_m}
\end{figure}

Fig.~\ref{fig:avrg_m} shows the mass of the final relic BH $m_\text{relic}$ at the time of the evaporation time as a function of the initial cluster size $N_i$ with three different values of $\omega$. The curves terminate at the maximal initial cluster size, beyond which the cluster would require longer BH domination to form. The input parameters are chosen as $m_f = 10^{7}\si{\gram}$ and $\beta=10^{-3}$. The relic mass with(out) accretion effects is shown by solid (dashed) lines. The curves are drawn by solving Eq.~\eqref{eq:diff_eq_M_N_orig} directly, and we find the analytical expression in Eq.~\eqref{eq:m_relic} tends to over-estimate the relic mass, especially at large $N_i$ (not shown in the figure). Depending on the equation of state prior to the BHD, the accretion effects can be $\order{1}$ on the initial BH mass for the merging. Accretion increases the relic mass by a similar factor. There is a ``knee'' in the curves for large $\omega$: it comes from the large scale suppression shown in Eq.~\eqref{eq:mu-definition}. The ``knee'' in the solid curves (with accretion) gets shifted towards lower $N_i$. This is due to the advancement of the equality time and correspondingly smaller horizon mass $M_H$. The suppression due to less time for structure formation then kicks in earlier. It is also interesting to note that the relic mass is lower for smaller $\omega$. In such cases, the BH requires more time to dominate the energy density, leaving less time for structure to form in which PBHs could merge. In any case, we see order-of-magnitude difference between the increase in the BH mass $m_\text{accret}/m_f$ and the initial cluster size $N_i$ in Fig.~\ref{fig:avrg_m}: the majority of BHs in clusters initially remains unmerged.

To compute the final BH mass spectrum, we need to use the PS mass function in Eq.~\eqref{eq:PS-mass-function}, where the cluster mass is related to the BH number in the cluster by $M_i = m_i N_i$. The BH mass spectrum at mass interval $(m, m+\dd{m})$ is given by the merged relic BHs ($m_\text{BH} = m_\text{relic}$) and unmerged BHs ($m_\text{BH}=m_\text{accret}$) \cite{holstClusteringRunawayMerging2024}
\begin{align}
    \dv{n_{\text{BH}}}{m_\text{BH}} (\teva) &\simeq \int_1^{N_{\text{max}}} \dd{N_i} \dv{n_\text{cl}(M_i, t)}{M} \delta(m - m_\text{relic}(M_i)) + {n}_{\text{BH}}(t) \delta(m-m_\text{accret}), \notag \\
                                            &= m_\text{relic} \dv{n_\text{cl} (M_i(m_{\text{relic}}), t)}{M}  \left| \pdv{m_\text{relic}}{N_i} \right|^{-1} + {n}_{\text{BH}}(t) \delta(m-m_\text{accret}),
                                  \label{eq:dndm_BH}
\end{align}
where the second term accounts for the unmerged BHs. Note that first integral start from unity. For very small clusters, the spherical collapse model for structure formation is not well justified. However, we note that this barely has any effects on the final constraint.

Due to merging, the PBH has a non-monochromatic mass function. This could potentially spoil the BBN constraint, if such heavy BHs Hawking evaporates during BBN. The constraint for monochromatic mass function is usually formulated in terms of the fraction of BH as dark matter \cite{carrNewCosmologicalConstraints2010a}
\begin{equation}
    f_{\text{BH}} \equiv \frac{\Omega_{\text{BH}}}{\Omega_\text{DM}}.
\end{equation}
In terms of an extended mass function, one defines the fraction per logarithmic interval \cite{carrPrimordialBlackHole2017, carrConstraintsPrimordialBlack2021}
\begin{equation}
    \psi (m_{\text{BH}}) \equiv \dv{f_{\text{BH}}}{\ln (m_\text{BH})} \equiv \frac{m_{\text{BH}}^2}{\rho_{\text{DM}}} \dv{n_{\text{BH}}}{m_{\text{BH}}}.
\end{equation}
The BH of interest here would have mostly evaporated away. However, the quantity can still be computed by \textit{not} taking the evaporation into account \cite{carrPrimordialBlackHoles2026}. Right before the evaporation of the BHs, the Universe's energy budget is taken almost entirely by the BHs $\Omega_\text{PBH}(\teva)\approx 1 $. This energy gets quickly transferred to radiation by Hawking radiation. If the BHs continue to exist after the evaporation time and PBH reheating takes place just as with Hawking evaporation, the current day BH energy parameter would be $\Omega_{\text{BH}}(a_0) = \Omega_\text{rad} (a_0) \Trh / T_0$ \cite{carrNewCosmologicalConstraints2010a}. Hence, the monochromatic part is typically very large \cite{carrNewCosmologicalConstraints2010a}:
\begin{equation}
    f_\text{BH} = \frac{\Omega_\text{rad}}{\Omega_\text{DM}} \frac{\Trh}{T_0} \simeq \lrfrac{\Trh}{\SI{1}{\eV}}.
    \label{eq:f_BH}
\end{equation}
Through Eq.~\eqref{eq:PS-mass-function} and first term of Eq.~\eqref{eq:dndm_BH}, the extended mass function can be expressed as 
\begin{equation}
    \psi(m) = f_\text{BH} \frac{m_\text{relic}}{N_i^2} \left|\pdv{m_\text{relic}}{N_i}\right|^{-1} \sqrt{\frac{2}{\pi}} \nu \exp(-\nu^2/2).
\end{equation}

\begin{figure}[ht]
    \centering
    \includegraphics[width=0.5\textwidth]{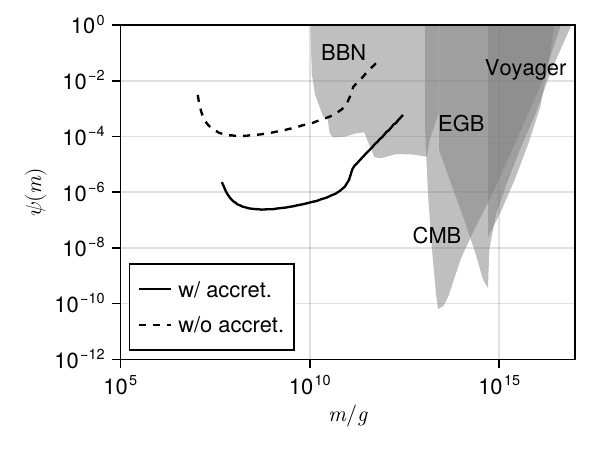}
    \caption{Extended mass spectrum along with the evaporation constraints. The two lines are computed using $m_f = 10^{7}\si{\gram}$, $\beta=10^{-3}$, and $\omega=1/3$. Grey shaded regions are current constraints taken from Ref.~\cite{carrPrimordialBlackHoles2026}.}
    \label{fig:df_dlnm}
\end{figure}

Fig.~\ref{fig:df_dlnm} shows the mass spectrum with(out) accretion as solid (dashed) lines, with $m_f = 10^{7}\si{\gram}$ and $\beta=10^{-3}$. The monochromatic mass functions in Eq.~\eqref{eq:f_BH} are much larger than unity and are not shown here. Accretion moves the mass spectrum to higher mass while reducing the fraction of DM in BH. The effects of accretion enter in two places: in the duration of BH domination in Eq.~\eqref{eq:PS-mass-function} (through the mass variance \eqref{eq:sigma-real}) and the reheating temperature $\Trh$ in Eq.~\eqref{eq:f_BH}. BHs become heavier due to accretion and leads to lower reheating temperature and lower (would-be) current energy density of the BHs. Hence, the mass function with the inclusion of accretion moves to higher mass and lower abundance.

To check the consistency of the resulting BH mass spectrum, one has to compare it with the constraints. The constraints are computed for monochromatic mass spectrum. For extended mass spectrum, one has to check the following criterion \cite{carrPrimordialBlackHole2017, carrConstraintsPrimordialBlack2021, bellomoPrimordialBlackHoles2018}
\begin{equation}
    \int \dd{\log(m)} \frac{\psi(m)}{f_{\text{max}}(m)} \leq 1,
    \label{eq:f_constraint}
\end{equation}
where $f_{\text{max}}(m)$ is the maximal allowed $f$ from observations in the mass interval. In Fig.~\ref{fig:df_dlnm}, the constraints from BBN, CMB, Extragalactic Background (EGB) and Voyage are shown in shaded regions (taken from Ref.~\cite{carrPrimordialBlackHoles2026}). Both solid and dashed curves are ruled out by Eq.~\eqref{eq:f_constraint}.

\begin{figure}[ht]
    \centering
    \includegraphics[width=0.7\textwidth]{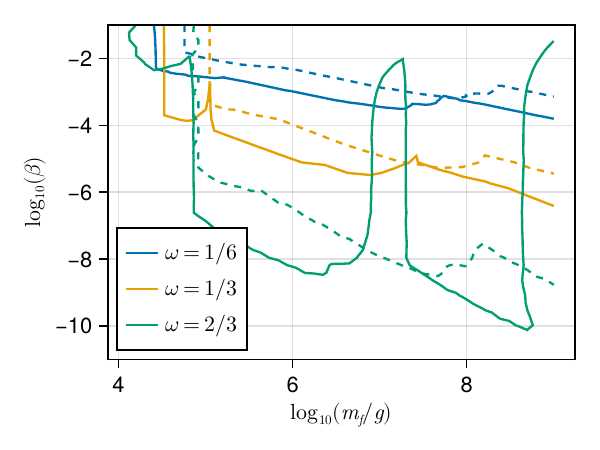}
    \caption{Merging constraint for PBH reheating. Regions above the curves are ruled out.}
    \label{fig:merging-constr-all}
\end{figure}

With Eq.~\eqref{eq:f_constraint}, one can check the compatibility of a certain scenario with $(m_f, \beta)$ with the observation constraints. The maximal $\beta$ allowed for the corresponding $m_f$ in terms of the evaporating BH constraints is in Fig.~\ref{fig:merging-constr-all}. The solid (dashed) lines are computed with(out) accretion effects. Note that formation mass $\gtrsim 10^{8} \si{\g}$ is ruled out by the PBH reheating condition (depending on accretion and equation of state $\omega$) and this is not shown in the plot. As has been noted before, the accretion makes the BH mass heavier and thus the BH mass function from merger would be shifted to larger masses. Since the evaporation constraints are for BH mass $\gtrsim 10^{10}\si{\gram}$, this mostly translated to pushing the lines towards lower $m_f$, as shown in the blue curve with $\omega=1/6$. However, there is another effect of accretion on the BH mass function: $\psi(m)$ gets lower as well. This is responsible for the ``features'' in the curves with $\omega=1/3$ and $2/3$. For certain formation masses, this reduction in the mass function can help with ``escaping'' the constraint. In particular, we see that for $\omega=2/3$, the region around $m_f \sim 10^7\si{\gram}$ is largely unconstrained: the resulting BH mass function with accretion effects lies just below the region for the BBN constraint. Thus, the inclusion of accretion might weaken the constraint for certain formation mass with high $\omega$. Note that beyond $m_f \gtrsim 10^{8} \si{\gram}$, the PBH reheating is constrained by the reheating temperature anyway, see Fig.~\ref{fig:BBN-cons}. Compared with Fig.~\ref{fig:BBN-cons}, we see that the merging constraints are not competitive with the BBN constraints on the isocurvature induced GWs.

\begin{figure}[ht]
    \centering
    \includegraphics[width=0.6\textwidth]{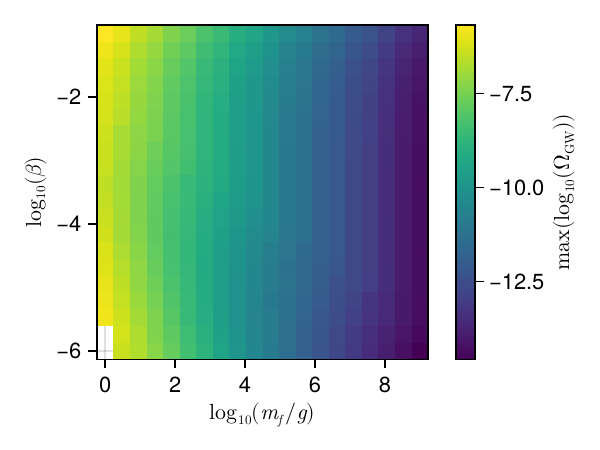}
    \caption{Upper bound on the $\Omega_\text{GW}$ from mergers right after the BH evaporation for $\omega=1/3$.}
    \label{fig:Omega-gw}
\end{figure}

GWs are also produced by the BH mergers. The mass loss fraction in merger events is $\simeq 0.05$ and is of similar magnitude as the BBN constraint. As only small fraction of BHs go through merger, the GWs of such origin pose weak constraint on the scenario, using the BBN $\Delta N_\text{eff}$ bound \cite{holstClusteringRunawayMerging2024}. Here, we want to check such statement quantitatively. 
% For a cluster of a given size: add an equation: 
% \begin{equation}
%     \dot{\rho}_\text{gw} + 4 H \rho_\text{gw} = \Delta\sum_{N_i} \Gamma_\text{merging}  m_\text{relic} n_\text{cl}(m_\text{relic})  \left|\pdv{m_\text{relic}}{N_i} \right|^{-1}  
% \end{equation}
% is the energy density of BH within the cluster of a certain size. I think this is unnecessary to compute, can think of some rough estimates. 
Firstly, we note that Eq.~\eqref{eq:diff_eq_M_N_orig} implies that merger \textit{always} takes place between two equal mass BHs. Then, we estimate the number of generations of merger with a relic mass $m_\text{relic}$ as 
\begin{equation}
    N_\text{merger}(N_i) = \frac{\log(m_\text{relic}(N_i)/m_f)}{\log(2-\Delta)}.
\end{equation}
Then the lost mass to GW is
\begin{equation}
    m_\text{lost}(N_i) = 2^{N_\text{merger}(N_i)}m_f - m_\text{relic}(N_i).
\end{equation}
If we assume this amount of mass/energy is released all at once at the formation of the final relic BH, just as a (over) optimistic estimate, the GW energy density can be written using Eq.~\eqref{eq:dndm_BH} as
\begin{equation}
    \rho_\text{GW}(a) < \int \dd{m_\text{BH}} m_\text{lost}(N_i) \dv{n_\text{BH}}{m_\text{BH}} \frac{a_\text{relic}(N_i)}{a},
\end{equation}
where $N_i=N_i(m_\text{BH})$ and the unmerged BHs in Eq.~\eqref{eq:dndm_BH} don't contribute. The last factor accounts for the dilution of GWs after production in matter domination and $a_\text{relic}$ is the redshift at relic BH formation. Thus, right at the evaporation, the fraction of energy density contains in the GWs can be (overly) estimated via Eqs.~\eqref{eq:PS-mass-function} and \eqref{eq:dndm_BH} as 
\begin{equation}
    % \Omega_{\text{GW, merger}}(\teva) < \frac{1}{ m_\text{accret}n_\text{BH} (\teva)} \int \dd{m_\text{BH}} m_\text{lost}(N_i)  n_\text{cl}(M_i(m_\text{relic}), \teva)  \left|\pdv{m_\text{relic}}{N_i} \right|^{-1} \frac{a_\text{relic}(N_i)}{a_\text{evap}}.
    \Omega_{\text{GW, merger}}(\teva) < \sqrt{\frac{2}{\pi}} \int \frac{\dd{N_i}}{N_i^2}   \nu \exp(-\nu^2/2) \frac{a_\text{relic}(N_i)}{a_\text{evap}}.
\end{equation}
Fig.~\ref{fig:Omega-gw} shows this upper bound on $\Omega_\text{GW, merger}$ for wide ranges of $(m_f, \beta)$ with $\omega=1/3$. We find the upper bound of $\Omega_\text{GW}$ to be several orders of magnitude lower than the current $\Delta N_\text{eff}$ bound for all relevant parameters $(m_f, \beta)$. And the actual $\Omega_\text{GW}$ could be order(s) of magnitude below our upper bound. Interestingly, the upper bound mostly depends on $m_f$ but not on $\beta$. The released mass/energy is proportional to the formation mass $m_f$, thus the strong dependence. On the other hand, $\beta$ has multiple effects on amount of GWs: it prolongs the BH domination via Eq.~\eqref{eq:N_eva}. It leads to more mergers but also more dilution for the produced GWs until radiation domination.  

In this Section, we find the accretion effects quite effective in shifting the extended mass function from mergers. This in turn can put bounds on the PBH reheating scenario. However, this is not competitive with the constraints from isocurvature induced GWs presented Section \ref{sec:iso-gw}. Before moving to the conclusion, we have a few remarks. The extended mass function obtained here is not likely to change the results of Section \ref{sec:iso-gw}: the majority of BHs remains unmerged, and the background evolution is still governed by the BHs of formation mass. We have discussed the merger dynamics using a few assumptions, one of which is the isolated cluster assumption. In the usual hierarchical cluster formation, large clusters are actually made of small clusters. The extended Press-Schechter formalism provides the probability of finding small clusters in a large cluster~\cite{moGalaxyFormationEvolution2010}. These small clusters most likely have gone through merger already and their mean BH mass is higher than the accreted/formation mass. Naively, one might expect heavier relic BH in the end. But properly studying this correction is out of the scope the current work, as the BH mass distribution most likely needs to be tracked.

\section{Conclusion} \label{sec:conc}
PBH reheating scenario enables an alternative explanation for the reheating of the Universe. PBHs could form from primordial fluctuations if these are large enough, $\delta \rho / \rho \gtrsim 10^{-2}$. These BHs can proceed to dominate the Universe with the right abundance and mass, as their energy density redshifts slower than that of the background fluid (usually taken as radiation). To recover the usual hot big bang, the energy density in BHs must be transferred to the radiation for successful BBN and this is done via Hawking evaporation. In this work, we focus on the effects of black hole accretion for such scenario. We find that the accretion not only changes the initial mass of the BHs, but also prolongs the black hole domination phase. In Section \ref{sec:background}, we have derived several analytical fits for the duration of each an era.

During the PBH domination phase, the density contrast grows linearly with the scale factor, $\delta \propto a$. After the sudden evaporation of the BHs to radiation, the gravitational potential $\Phi$ starts to oscillate due to non-negligible radiation pressure. This can generate a significant amount of GWs, possibly breaching the BBN $\Delta N_\text{eff}$ bound. This in turn puts constraints on the initial abundance $\beta$ and the formation mass $m_f$, discussed in Section \ref{sec:iso-gw}. With the inclusion of accretion, we find that such constraints are shifted towards lower initial abundance $\beta$ and formation mass $m_f$. The allowed maximal $m_f$ is reduced, since accretion increases the BH mass and thus decreases the reheating temperature from its Hawking radiation. We also find that such shift in constraint is more dramatic for large equation of state parameter of the background fluid $\omega$, where the accretion effect is more pronounced. With the linear growth of the density contrast, clusters of BHs could form using the standard Press-Schecter formalism as discussed in Section \ref{sec:merger}. Inside the clusters, merger could take place and heavier BHs can be produced. This gives the scenario a mechanism to produce (much) heavier BHs from the initial BHs. BHs around $10^{10}\si{\gram}$ to $10^{17}\si{\gram}$ are constrained by various observations. Thus, the merging could put additional constraints on the PBH scenario. Since the accretion changes the duration of PBH domination and the reheating temperature, it also affects such constraints. We find that the inclusion of accretion leads to a moderately more strict constraint, albeit still not very competitive compared to the constraint coming from isocurvature induced gravitational waves.

In this work, we have assumed a monochromatic BH mass function at formation. Depending on the width of the peak of the primordial power spectrum responsible for PBH formation, this can be inadequate. With an extended mass spectrum at formation, the transition from matter domination to radiation domination would take longer and the instantaneous approximation is not valid anymore. Ref.~\cite{inomataGravitationalWaveProduction2020} has checked the effects of such a mass function and found that the peak amplitude of GWs could reduce by a couple of orders of magnitude. This, however, would possibly make the merging constraint stronger than the isocurvature induced GWs constraint. In the clustering and merging treatment in Section \ref{sec:merger}, only the mean BH mass in a cluster is tracked, instead of the full distribution. This can be improved by using coagulation equation, see e.g. Ref.~\cite{doctorBlackHoleCoagulation2020}. The accretion dynamics in this work is based on Ref.~\cite{dasImpactGeneralRelativistic2025}. We note that the mass change rate due to the accretion shown in Eq.~\eqref{eq:Gamma-accret} is quite general $\dot{m} \propto \rho_\infty m^2$. We still expect similar results using different accretion models. Another constraint on PBH reheating can be realised via (over)production of dark matter. However, it requires knowledge of the nature of such dark matter and is out of scope of this work, see e.g.~\cite{dasImpactGeneralRelativistic2025, bernalSuperradiantProductionHeavy2022}. It might also be interesting to have a complete model for inflation, from which the primordial curvature perturbation at all scales and BH formation dynamics can be accurately captured. The GWs from adiabatic initial conditions can then be computed as well.

\acknowledgments{
    We are grateful to Manuel Drees for valuable discussions and comments on the draft. We thank Yong Xu for comments.
}

\bibliographystyle{JHEP}
\bibliography{refs.bib}

% \printbibliography
\end{document}